# DFT insights into the new Hf-based chalcogenide MAX phase Hf$_2$SeC

M. A. Ali[a,*] Muhammad Waqas Qureshi[b, c, **]

[a]Department of Physics, Chittagong University of Engineering and Technology (CUET), Chattogram-4349, Bangladesh
[b]State Key Laboratory of Advanced Welding and Joining, Harbin Institute of Technology, Harbin 150001, China
[c]School of Materials Science & Engineering, Harbin Institute of Technology, Harbin 150001, China

**Abstract**

The physical characteristics of the novel chalcogenide MAX phase Hf$_2$SeC have been investigated using the DFT method. The obtained lattice constant and elastic constants ($C_{ij}$) are compared with previous results to check the consistency of our setting parameters during calculations Moreover, the elastic properties such as elastic constants, moduli, anisotropy, and hardness are also compared with preexisting MAX phases of its kind. The checking of mechanical stability has been done based on $C_{ij}$ in accordance with the previously stated stability criteria. The reason for the higher hardness of Hf$_2$SeC compared to Hf$_2$SC is explained using the density of states (DOS). The brittleness character of Hf$_2$SeC has been revealed using the Pugh ratio, Poisson's ratio, and Cauchy pressure. The electronic properties (band structure and charge density mapping) of Hf$_2$SeC are studied to disclose the metallic nature as well as bonding nature within the titled chalcogenide. The anisotropy in both electronic conductivity and mechanical properties is investigated. The temperature and pressure dependence of volume, Grüneisen parameter ($\gamma$), Debye temperature ($\Theta_D$), thermal expansion coefficient (*TEC*), and specific heat at constant volume ($C_v$) are explored. In addition, minimum thermal conductivity ($K_{min}$) and melting point ($T_m$) are studied to explore its suitability for high-temperature applications.

## 1. Introduction

Recently, chalcogenide MAX phase members have been extended by the synthesis of two new chalcogens (Se) containing MAX phase: M$_2$SeC (M = Hf, Zr) [1,2]. The scientific community is trying to expand the family member of MAX phase materials for a long time, motivated by their peculiar properties, as a result, more than 150 MAX phases have already been revealed [3]. To do this, scientists have focused on the tuning of the compositions and structure to accomplish

Email addresses: ashrafphy31@cuet.ac.bd, waqasmse@hit.edu.cn

superior properties. The recent attempts reported so far as new compounds [1,2,4–11] solid solutions [12–21]. Some other MAX phases have also been reported with slightly different structures: $M_2A_2X$ [22–24] and $M_3A_2X$ [25], rare-earth i-MAX phases [26,27], 212 MAX phases [28–31], 314 MAX phases [28,32,33]. The most recent extension of X element up to B has also been reported as MAX phase borides [5,6,34–39].

Now, the question is what aspect of the MAX phase triggers the scientists. The prime feature of the MAX phase that attracts mainly is the unification of metallic and ceramics properties. Again, the question is how it is possible to combine both characteristics. The answer lies within the structure of the MAX compounds in which there are M-A bonds responsible for metallic character and covalent M-X bonds that contribute to the ceramics character [40,41]; consequently, the conductivity (electrical and thermal), the strength of mechanical character, machinability are like metals, and the high-temperature mechanical parameters are good, corrosion and oxidation resistance also good as like as ceramics materials [42] which are exhibited by the MAX phases. Owing to the aforesaid characteristics, their potential applications include a very long list that can be found elsewhere [3,21,42–44].

The S containing MAX phases [$M_2SC$ (M=Ti, Zr, Hf, Nb)] have been studied both experimentally and theoretically [5,6,35,45–50]. Bouhemadou et al [45] studied the elastic and electronic properties of $M_2SC$ (M=Ti, Zr, Hf) phases. Superconducting phase $Nb_2SC$ has been reported by Shein et al [46]. A comprehensive survey of elastic properties of the $M_2AX$ phase was carried out by Cover et al [50] where S containing phases were included. $Zr_2SC$ MAX compound was investigated by Feng et al in which the elastic and electronic properties were considered. Nasir et al [48,49] have selected Zr and Nb-based S containing phases for their DFT-based study. S-based MAX phase borides were synthesized by Rackl et al [5,6] which were further investigated [35]. To our knowledge, only the $Zr_2SeC$ compound is studied experimentally [1] and theoretically [51]. Wang et al [2] synthesized $Hf_2SeC$ and studied its thermal expansion coefficient and stiffness constants ($C_{ij}$) only. The obtained results were compared with those of $Hf_2SC$, $Zr_2SeC$, and $Zr_2SC$, $Ti_2SC$ MAX phases where available. This too little information is not adequate to disclose the potential of $Hf_2SeC$ for industrial use in various areas, in place of other MAX phases that are already been used. For example, the mechanical behavior exploring parameters such as elastic moduli, ductile/brittleness behavior, mechanical hardness parameters are pre-requisite for the prediction of the applications as structural components at high-temperature [3,52,53]. Knowledge of Debye temperature, minimum thermal conductivity, Grüneisen parameter as well as melting temperature should be known to predict its high-temperature applications [54,55]. Calculation of electronic band structure, the density of states, and charge density mapping is very useful to disclose the bonding nature clearly and explain the mechanical and thermal properties. Thus, a comparative study of newly synthesized chalcogenide could be of scientific interest.

Therefore, we planned to perform a DFT-based investigation of the synthesized $Hf_2SeC$ and the structural, elastic, electronic, and thermodynamic properties of $Hf_2SeC$ are presented in this paper. To be more acceptable, the obtained results were compared with those of $Hf_2SC$, $Zr_2SeC$, $Zr_2SC$, and $Nb_2SC$ MAX phases, where available.

## 2. Computational methodology

The results presented here are calculated using the Cambridge Serial Total Energy Package (CASTEP) code [56,57] based on the density functional theory. The generalized gradient approximation (GGA) of the Perdew–Burke–Ernzerhof (PBE) [58] type of functional was used as exchange and correlation term. The C - $2s^2\ 2p^2$, Se- $4s^2\ 4p^4$, and Hf - $5d^2\ 6s^2$ electronic orbitals were considered for pseudo-potential calculations. A k-point [59] mesh of size $12 \times 12 \times 4$ and the cutoff energy of 500 eV was used for the calculations. Further details of the calculations can be found elsewhere [51].

## 3. Results and discussion
### 3.1 Structural properties

The crystal structure of $Hf_2SeC$ is presented in Fig. 1, it belongs to the hexagonal system and the space group is $P6_3/mmc$ (194) [42]. The unit cell is composed of eight atoms, hence, its chemical formula revealed the two formula units. The octahedron $Hf_6C$ is situated in between two atomic Se layers that can be shown in ref. [51]. The Wyckoff positions of M (Hf), A (Se), X (C) atoms are 4f (1/3, 2/3, $Z_M$), 2a (1/3, 2/3, 3/4) and 2c (0, 0, 0), respectively [60]. The physical properties of $Hf_2SeC$ were further calculated after the geometrical relaxation of the unit cell. The cell parameters of the relaxed structure are given along with the previous experimental and theoretical results[2] to assure the setting of the parameters used for calculations. For example, the obtained value of *a* is just 0.47% and *c* is just 0.88% larger than the experimental results. The discrepancy between the calculated values (present and previous) is much smaller (0.06% and 0.39% for *a* and *c*, respectively) compared to the experimental values.

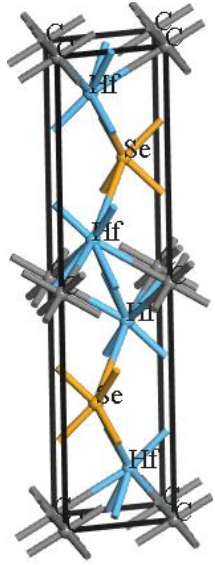

Fig. 1. Crystal structure (unit cell) of the $Hf_2SeC$ compound.

Table 1. Calculated lattice parameters (*a* and *c*), *c/a* ratio, and atomic positions of $Hf_2SeC$ MAX phase.

| *a* (Å) | *c* (Å) | c/a | *Ref.* | Positions | Zr | Se | C |
|---|---|---|---|---|---|---|---|
| 3.438 | 12.501 | 3.636 | *This study* | x | 1/3 | 1/3 | 0 |
| 3.422 | 12.391 | 3.621 | *Expt.* [2] | y | 2/3 | 2/3 | 0 |
| 3.436 | 12.452 | 3.624 | *Theo.* [2] | z | 0.0945 0.0943 [2] | 3/4 | 0 |

### 3.3 Mechanical properties

Wang et al [2] calculated the stiffness constants only with an intention to correlate these with the thermal expansion coefficient of MAX phases. However, other important parameters like elastic moduli, failure mode based on Pugh/Poisson's ratio, mechanical hardness parameters, elastic anisotropic indices, etc. are remained unexplored and will be focused on in this section. To do this, we have re-calculated the stiffness constants ($C_{ij}$) of $Hf_2SeC$ using the strain-stress method [30,32,61–63] and then used them further to investigate the other parameters mentioned above. The five independent stiffness constants of $Hf_2SeC$ are given in Table 2 along with stiffness constants of other S and Se-based MAX phases $M_2AC$ (M = Hf, Zr, Nb, A = S, Se), where available. Though, $Hf_2SeC$ is experimentally realized, the mechanical stability of $Hf_2SeC$ is also checked by use of stiffness constants ($C_{ij}$) based on the Max Born [64] and Mouhat *et al.*[65] stability conditions as given here: $C_{11} > 0$, $C_{11} > C_{12}$, $C_{44} > 0$, $(C_{11} + C_{12})C_{33} - 2(C_{13})^2 > 0$. Since it ($Hf_2SeC$) is experimentally synthesized but for realistic use, checking of mechanical stability is mandatory. Table 2 confirms that the obtained $C_{ij}$s meet the stability conditions; hence, there is no doubt about the mechanical stability. In addition, $C_{ij}$s also provide knowledge regarding bonding strength along different axes, hardness, mechanical anisotropy, etc. For instance, the values $C_{11}$ and $C_{33}$ represent the stiffness along [100] and [001] directions. As evident from Table 2, $C_{11}$ is smaller than $C_{33}$ for all considered chalcogenide MAX phases implying lower deformation resistance along the *a*-axis than the *c*-axis. Moreover, $C_{44}$ is lower than $C_{33}$ and $C_{11}$, certifying an easy shear deformation compared to that of along the axial directions. Finally, the values of $C_{11}$, $C_{33}$, and $C_{44}$ ($C_{11} \neq C_{33} \neq C_{44}$) confirmed the uneven bonding strength along with the *a* and *c*-axis as well as shear direction which is owing to the differences in the arrangement of atoms in different directions.

Table 2. The elastic constants, $C_{ij}$ (GPa), bulk modulus, $B$ (GPa), shear modulus, $G$ (GPa), Young's modulus, $Y$ (GPa), Hardness parameters, $H_{Chen}$(GPa) and $H_{Miao}$(GPa), Pugh ratio, $G/B$, Poisson ratio, $v$ and Cauchy Pressure, $CP$ (GPa) of $Hf_2SeC$, along with those of $Zr_2SeC$ and $M_2SC$ (M = Zr, Hf, Nb).

| Phase | $C_{11}$ | $C_{12}$ | $C_{13}$ | $C_{33}$ | $C_{44}$ | $B$ | $G$ | $Y$ | $H_{Chen}$ | $H_{Miao}$ | $G/B$ | $v$ | $CP$ | Reference |
|---|---|---|---|---|---|---|---|---|---|---|---|---|---|---|
| $Hf_2SeC$ | 287 | 77 | 98 | 306 | 130 | 158 | 113 | 273 | 18.47 | 21.69 | 0.72 | 0.21 | -53 | This study |
|  | 308 | 95 | 105 | 314 | 135 |  |  |  |  |  |  |  |  | [2] |
| $Hf_2SC$ | 311 | 97 | 121 | 327 | 149 | 181 | 120 | 295 | 17.35 | 21.72 | 0.66 | 0.23 | -52 | [35] |
| $Zr_2SeC$ | 260 | 97 | 96 | 293 | 128 | 154 | 100 | 247 | 14.85 | 17.79 | 0.65 | 0.23 | -30 | [51] |
| $Zr_2SC$ | 295 | 89 | 102 | 315 | 138 | 166 | 115 | 280 | 17.89 | 21.57 | 0.69 | 0.22 | -49 | [35] |
| $Nb_2SC$ | 316 | 108 | 151 | 325 | 124 | 197 | 105 | 267 | 11.58 | 15.84 | 0.53 | 0.27 | -16 | [35] |

The stiffness constants were then used to compute the Bulk modulus ($B$) and shear modulus ($G$) using Hill's approximation [66] which is the average of Voigt [67] and Reuss [68] models as follows:

$B = (B_V + B_R)/2$, where $B_V = [2(C_{11} + C_{12}) + C_{33} + 4C_{13}]/9$ and

$B_R = C^2/M$; $C^2 = C_{11} + C_{12})C_{33} - 2C_{13}^2$; $M = C_{11} + C_{12} + 2C_{33} - 4C_{13}$. $B_V$ indicates the largest value of $B$ (Voigt bulk modulus) and $B_R$ express the lowest value of $B$ (Reuss bulk modulus). Similarly, $G$ was computed using the following equations:

$G = (G_V + G_R)/2$, where $G_V = [M + 12C_{44} + 12C_{66}]/30$ and

$G_R = \left(\frac{5}{2}\right) [C^2 C_{44} C_{66}]/[3B_V C_{44} C_{66} + C^2(C_{44} + C_{66})]$; $C_{66} = (C_{11} - C_{12})/2$.

After then, the $B$ and $G$ were used to calculate the Young's modulus ($Y$) and Poisson's ratio ($v$) using the following equations[35] : $Y = 9BG/(3B + G)$ and $v = (3B - Y)/(6B)$.

Finally, $B$, $G$, $Y$ and $v$ were used to calculate the Hardness parameters using the equations [69,70]: $H_{Chen} = 2(k^2 G)^{0.585} - 3$ (where, Pugh's ratio, $k=G/B$) and $H_{Miao} = \frac{(1-2v)E}{6(1+v)}$.

Now, let us have a comparative discussion based on the values presented in Table 2. The value of $C_{11}$ for Hf$_2$SeC is higher than that of another Se containing MAX phase, Zr$_2$SeC. On other hand, it is lower than those other S-based MAX phases, which is also true for $C_{33}$. Next, $C_{44}$ of Hf$_2$SeC is higher than that of Zr$_2$SeC and Nb$_2$SC but lower than that M$_2$SC (M = Hf, Zr). As it is well known, $C_{44}$ is a very good hardness predictor, i.e., more directly related to the elastic moduli compared to other stiffness constants and also used for the measurement of deformation resistance. Thus, according to the $C_{44}$, the compounds can be ranked as follows: Hf$_2$SC> Zr$_2$SC >Hf$_2$SeC> Zr$_2$SeC > Nb$_2$SC. Among the elastic moduli, $G$ is a more accurate predictor for shear deformation compared to $B$ and/or $Y$. The ranking of the compounds is same as for $C_{44}$, i.e., Hf$_2$SC> Zr$_2$SC > Hf$_2$SeC > Zr$_2$SeC > Nb$_2$SC. The $Y$ is a good predictor for the stiffness measurement of the solids and the solid is said to be stiffer with a higher $Y$ value. Thus, the compounds can be ranked as follows: Hf$_2$SC> Zr$_2$SC >Hf$_2$SeC> Nb$_2$SC > Zr$_2$SeC. However, the $B$ of Nb$_2$SC somehow has the highest value. The ranking based on the $B$ is also the same for other compounds except for Nb$_2$SC, i.e., Hf$_2$SC > Zr$_2$SC > Hf$_2$SeC > Zr$_2$SeC. Next, the hardness parameters using the Chen's [69] and Miao [70] formulae were calculated and presented in Table 2. As shown, The $H_{Chen}$ of Hf$_2$SeC is comparatively higher than other chalcogenides presented here despite its lower values of elastic constants compared to Hf$_2$SC and Zr$_2$SC. The elastic constants don't reflect material's hardness directly, though, the parameters have usually higher values for harder solids. Thus, the titled chalcogenide is expected to be harder than the others presented here. $H_{miao}$ of Hf$_2$SeC is also higher than that of other chalcogenides except for Hf$_2$SC where $H_{miao}$ of Hf$_2$SeC is 0.13% lowered than that of Hf$_2$SC. In addition, Pan et al [71] demonstrated that the Pugh ratio (which is also used in Chen's formula) is indirectly related to the Vickers hardness, which is highest for $Hf_2SeC$ compared to other compounds considered here. It is noted that the hardness parameters of $Hf_2SeC$ are lowered than that of $Ti_2SC$ [32] ($H_{Chen}$- 23.48 GPa, $H_{miao}$-29.85 GPa, the elastic moduli are also higher) chalcogenide MAX phase. Thus, by taking into consideration the above facts, it is obvious to say that the titled compound possesses the 2$^{nd}$ highest hardness among the chalcogenide MAX phases. Despite the explanation stated above, still, the mystery (hardness of $Hf_2SC$ is lower than that of $Hf_2SeC$ though the stiffness constants and elastic moduli are higher) should be solved by further clarification in terms of the total density of states (DOS).

To check the ductility/brittleness character we used three well formalisms. Pugh ratio ($G/B$), the ratio of bulk to shear modulus, is a well-accepted parameter to separate the ductile and brittle materials with a critical value of 0.571. A value of $G/B$ greater than 0.571 implies the brittleness character of solids and vice versa [72]. As evident, the titled MAX phase is brittle like its precursor $Hf_2SC$. The Poisson's ratio [73] is another parameter used to define a solid either ductile or brittle with a critical value of 0.26. Values lower than 0.26 indicate the brittleness character of solids and vice versa. It is seen in Table 2, that the titled phase is brittle again. Next, the Cauchy pressure [74] is also used to define a material either ductile or brittle. Values of a negative sign indicate the brittle solids while the positive sign indicates the ionic ductile solids. Table 2 revealed that the titled Se based MAX phase is brittle. A large negative value also indicates the dominant nature of a covalent bond in which a mixture of bonding is present. Thus, the covalent bond is dominant in the $Hf_2SeC$ MAX phase where a mixture of covalent and ionic bonding is present.

### 3.3 Electronic properties

Fig. 2 (a) shows the DOS of $Hf_2SeC$ and $Hf_2SC$ in which black vertical lines indicate the peaks in the DOS of $Hf_2SeC$ and red lines indicate the peaks in the DOS of $Hf_2SC$. The reference lines clearly indicate the peak shifting to the left when we go from S to Se. The peaks in the DOS are the results of hybridization among the different states involved in solids. The strength of the bonding between states depends on the position. The peak in the low energy sides, higher the bonding strength[19,35,75]. Thus, the bonding among the states in $Hf_2SeC$ is expected to be stronger compared to Hf$_2$SC as a consequence, the hardness of Hf$_2$SeC is also expected to be higher compared to Hf$_2$SC that is found in the present study. In addition, the DOS at the Fermi level is around 1.33 states/eV/cell for both phases, which implies the metallic nature of Hf$_2$SeC. To confirm electronic conductivity, we have calculated the electronic band structure presented in the following section.

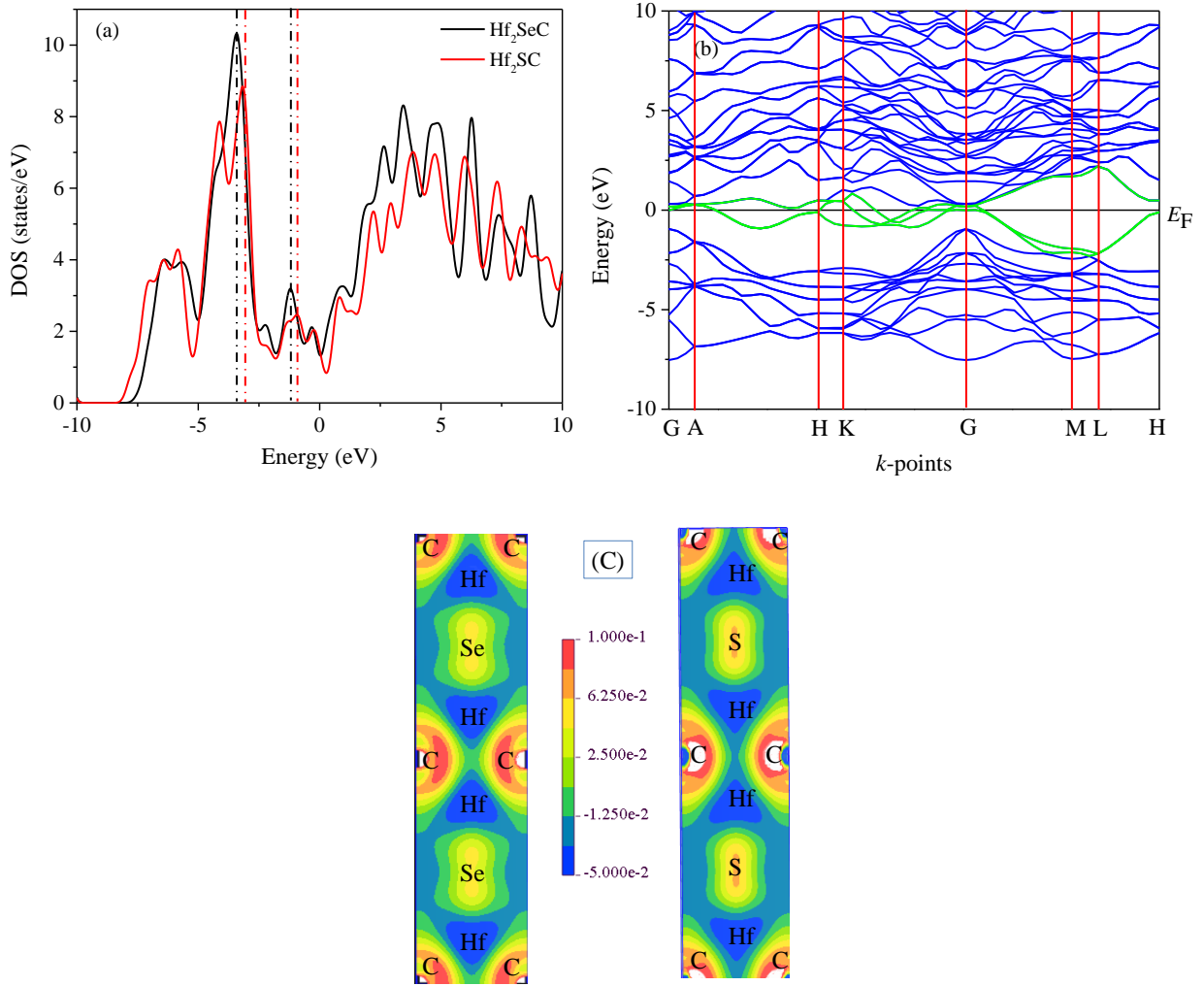

Fig. 1 (a) The DOS of Hf$_2$AC (A = Se, S), (b) electronic band structure of Hf$_2$SeC and (c) charge density mapping of Hf$_2$AC (A = Se, S).

The electronic band structure is also an important result concerning its electronic conductivity. The electronic band structure of Hf$_2$SeC calculated from the relaxed structure is shown in Fig. 2 (b) along the high symmetry lines: Γ–A–H–K–Γ–M–L–H. Like all other MAX phases[76], Hf$_2$SeC also exhibits metallic nature that can be confirmed from the band structure in which overlapping of valence and conduction band is observed. Another interesting character is also noticed from the band structure, the energy dispersion along (*c*-direction): Γ–A, K–H, and L–M directions are smaller compared to the energy dispersion (basal plane): H–K, Γ–M, and L–H, thus, the electronic conductivity is not equal in the basal plane and c-direction. Smaller energy dispersion is owing to the higher electronic effective mass tensor that leads to a smaller conductivity and vice versa [19]. Thus, the typical conductivity nature for MAX phases [19,30,77] is also exhibited by $Hf_2SeC$: higher conductivity is observed in the basal plane in comparison with that of *c*-direction.

Fig. 2 (c) displays the CDM (in the units of e/Å$^3$) in the (100) crystallographic plane which is designed to explore the nature of the bonding among the different atoms for $Hf_2SeC$ and $Hf_2SC$. The positions of the atoms are denoted in the figure. As evident, the Hf-C boning is a purely covalent bond like other MAX phases. The Hf-Se/S bonding is a mixture of a covalent and ionic type which is also another common nature for MAX phase materials. The replacement of A atom from Se to S results in a considerable change in the CDMs. As seen, the charge density (CD) at A (S) position increases. Thus, the common bonding nature of MAX phases is revealed by CDM for both $Hf_2AC$ (A = Se, S) which is responsible for the combined properties of metals and ceramics.

## 3.4 Mechanical anisotropy

Because of the layered hexagonal symmetry for the MAX phase materials, the atomic arrangement is different along *a* and *c*-direction, consequently, unequal values of the $C_{11}$ and $C_{33}$ are obtained (Table 2). As a further consequence, mechanical anisotropy is observed like other MAX phases [30,32,78–80]. The important phenomena such as microcracks and anisotropic plastic deformation are the results of the mechanical anisotropy that shows a significant contribution in realizing the mechanical stability of the solids in extreme conditions. Due to the aforementioned reasons, the anisotropic nature of the mechanical properties characterizing parameters has been studied. To do this, We have calculated the different anisotropic factors for the {100}, {010} and {001} directions calculated using the equations[81]:

$$A_1 = \frac{1/6(C_{11}+C_{12}+2C_{33}-4C_{13})}{C_{44}}, A_2 = \frac{2C_{44}}{C_{11}-C_{12}}, A_3 = A_1 \cdot A_2 = \frac{1/3(C_{11}+C_{12}+2C_{33}-4C_{13})}{C_{11}-C_{12}}$$

The obtained values are presented in Table 3 along with other chalcogenide MAX phase carbides that confirm the anisotropy of Hf$_2$SeC where the non-zero values of $A_i$'s indicating the level of anisotropy. The B along $a$ and $c$-direction are estimated using the relations $B_a = a\frac{dP}{da} = \frac{\Lambda}{2+\alpha}$ and $B_c = c\frac{dP}{dc} = \frac{B_a}{\alpha}$, respectively [82] where $\Lambda = 2(C_{11} + C_{12}) + 4C_{13}\alpha + C_{33}\alpha^2$ and $\alpha = \frac{(C_{11}+C_{12})-2C_{13}}{C_{33}-C_{13}}$. As evident from Table 3, $B_a \neq B_c$ means the B of Hf$_2$SeC is anisotropic in nature. The linear compressibility ($k$) along the $a$ and $c$-axis is estimated by the equation [83]:

$$\frac{k_c}{k_a} = C_{11} + C_{12} - 2C_{13}/(C_{33} - C_{13})$$

The ratio of $k_c/k_a$ is not one (1) ($k_c/k_a$ is equal to 1 for isotropic materials) that confirms the anisotropic nature of Hf$_2$SeC. The universal anisotropic index $A^U$ is estimated by the following equation [84]:

$$A^U = 5\frac{G_V}{G_R} + \frac{B_V}{B_R} - 6 \geq 0 \tag{15}$$

where B and G are obtained by Voigt and Reuss models. Since, the values of $A^U$ are greater than zero, indicating the anisotropic nature of Hf$_2$SeC. At end of this section, it is seen that the titled chalcogenide exhibits anisotropy in its elastic properties like other chalcogenide MAX phase carbides (Table 3).

Table 3. The anisotropic factors, $A_1$, $A_2$, $A_3$, $B_a$, $B_c$, $k_c/k_a$, and universal anisotropic index $A^U$ of M$_2$AC (M=Hf, Zr; A = Se, S) and Nb$_2$SC.

| Phase | $A_1$ | $A_2$ | $A_3$ | $k_c/k_a$ | $B_a$ | $B_c$ | $A^U$ | Ref. |
|---|---|---|---|---|---|---|---|---|
| Hf$_2$SeC | 0.75 | 1.24 | 0.93 | 0.81 | 443 | 548 | 0.081 | This |
| Hf$_2$SC | 0.65 | 1.39 | 0.90 | 0.81 | 505 | 627 | 0.178 | [35] |
| Zr$_2$SeC | 0.73 | 1.57 | 1.14 | 0.84 | 383 | 904 | 0.218 | [51] |
| Zr$_2$SC | 0.73 | 1.34 | 0.98 | 0.85 | 412 | 955 | 0.113 | [35] |
| Nb$_2$SC | 0.63 | 1.19 | 0.75 | 0.70 | 530 | 756 | 0.161 | [35] |

In order to better comprehend the mechanical anisotropy, the directional dependent 2D projections and 3D plots of elastic properties are used to illustrate this anisotropic behavior. Fig. 3 depicts the demonstration of anisotropy in Young's and shear moduli along with linear compressibility ($K$) and Poisson's ratio ($v$). The Y and K are isotropic in the $xy$ plane while anisotropic in the $xz$ and yz planes (Fig. 3 a and b), which are confirmed by perfectly circular ($xy$ plane) and non-circular ($xz$ and $yz$ planes) 2D projections. In the $xy$ plane, the Y and K values are equal at the *x* and *y* axis. In the *xz* and *yz* planes, the $Y_{max}$ is at the 45° of horizontal and vertical axis whereas $K_{max}$ is at the horizontal axis (Table 4).

Table 4. Maximum and minimum values of Youngs modulus, Y (GPa), linear compressibility, K (TPa$^{-1}$), shear modulus, G (GPa), and Poisson's ratio, υ (GPa) of M$_2$AC (M = Hf, Zr; A = Se, S) and Nb$_2$SC.

| Phases | $Y_{min}$ | $Y_{max}$ | $A_Y$ | $K_{min}$ | $K_{max}$ | $A_K$ | $G_{min}$ | $G_{max}$ | $A_G$ | $υ_{min}$ | $υ_{max}$ | $Aυ$ | Ref. |
|---|---|---|---|---|---|---|---|---|---|---|---|---|---|
| Hf$_2$SeC | 247.7 | 295.4 | 1.19 | 1.831 | 2.255 | 1.23 | 99.07 | 130.27 | 1.31 | 0.133 | 0.277 | 2.07 | This |
| Hf$_2$SC | 255.03 | 327.56 | 1.28 | 1.59 | 1.97 | 1.24 | 98.48 | 148.95 | 1.51 | 0.099 | 0.327 | 3.28 | [35] |
| Zr$_2$SC | 250.08 | 306.39 | 1.22 | 1.78 | 2.12 | 1.19 | 101.19 | 137.27 | 1.35 | 0.110 | 0.290 | 2.53 | [35] |
| Zr$_2$SeC | 208.93 | 277.50 | 1.32 | 1.913 | 2.288 | 1.19 | 81.071 | 127.59 | 1.57 | 0.078 | 0.343 | 4.394 | [51] |
| Nb$_2$SC | 217.27 | 297.02 | 1.36 | 1.32 | 1.88 | 1.43 | 88.96 | 128.57 | 1.50 | 0.154 | 0.393 | 2.55 | [35] |

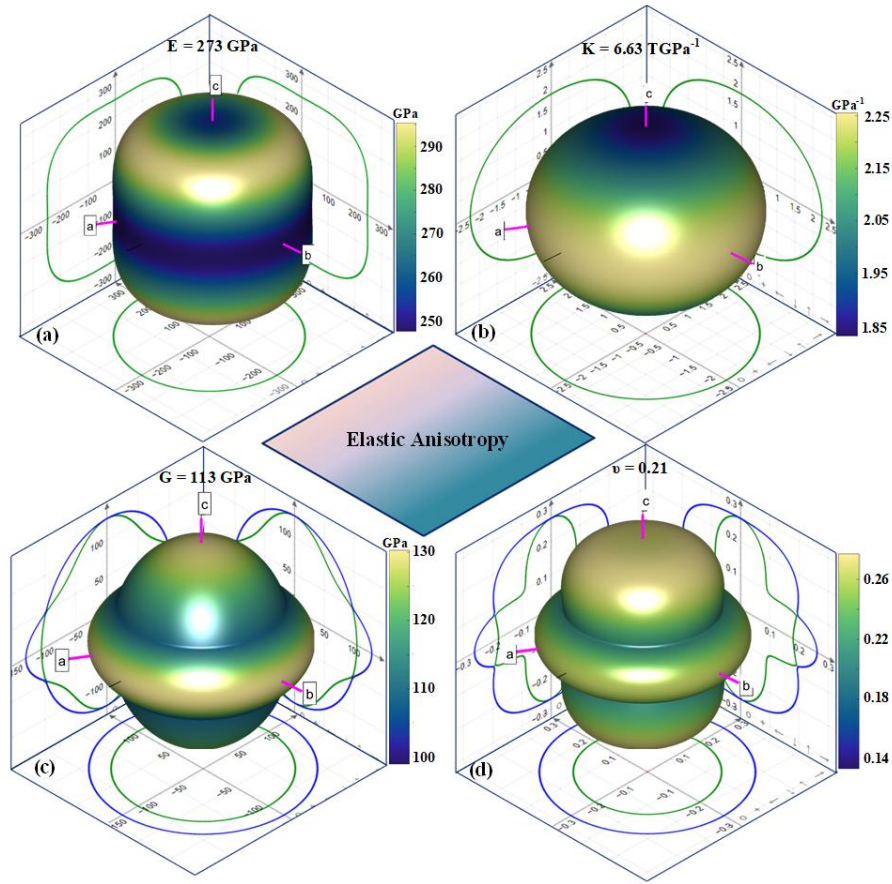

Fig. 3. The 2D projections and 3D plots of (a) Young's modulus, (b) linear compressibility, (c) shear modulus, and (d) Poisson's ratio in Hf$_2$SeC.

The color contrast on their 3D plots is consistent with the values in the 2D projections. The 2D projections of G and $\upsilon$ exhibit inner and outer surfaces (represented by green and blue lines) shown in Fig. 3 (c and d). The circular 2D projections of the inner and outer surface in *xy* plane reveal their isotropic nature similar to *Y* and *K*. In *xz* and *yz* planes, the outer surface of *G* and *υ* has the maximum value ($G_{max}$ and $\upsilon_{max}$) at the horizontal axis while the inner surface has a maximum value ($G_{max}$ and $\upsilon_{max}$) at the vertical axis. Due to this, the color contrast in 3D plots shows the maximum gradient at the *x*, *y* and *z*-axis while the minimum gradient at the 45º of the horizontal and vertical axis in the *xz* and *yz* planes. These computed values of *Y*, *G*, *K*, and *υ* are tabulated along with their anisotropic index in Table 4. This analysis shows that elastic properties are non-identical in all of the crystallographic planes and the degree of anisotropy is maximum along the *z*-axis. In comparison with other chalcogenide MAX phases, the level of anisotropy for $Hf_2SeC$ is lowest [Table 4].

## 3.5. Thermodynamic properties

MAX phases exhibit promising high-temperature applications such as heating elements, protective coatings, electrical contacts, and have the potential for thermal barrier coatings (TBCs) [85]. It is important to disclose their thermodynamic properties for possible applications. Therefore, we used the quasi-harmonic Debye model [86,87] to compute the thermodynamic properties of $Hf_2SeC$ in the temperature and pressure range of 0 to 1000 °C and 0 to 50 GPa, respectively. The relationship of unit cell volume as a function of temperature and pressure is presented in Fig. 4. With the increase in temperature and pressure, the unit cell volume expands and contracts linearly. The expansion rate as a function of temperature is relatively less at 50 GPa as compared to at 0 GPa. Here, the calculated unit cell volume of 128.3 Å$^3$ at 0 GPa and 300 K, is consistent with the equilibrium volume computed during optimization. The Grüneisen parameter (γ) for $Hf_2SeC$ is following the same trend as unit cell volume. There is a linear increase and decrease in the Grüneisen parameter with an increase in temperature and pressure, respectively as shown in Fig. 5.

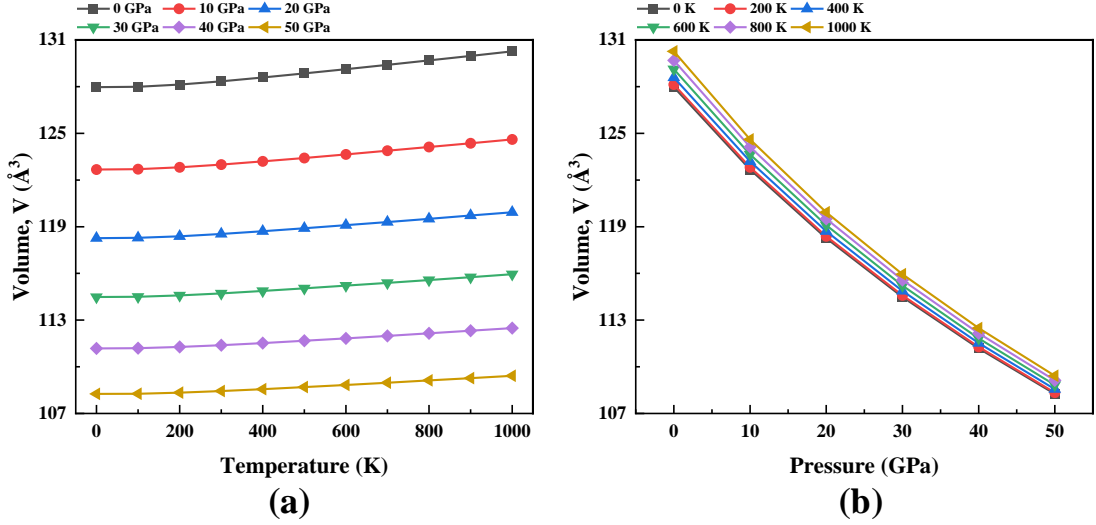

Fig. 4. Variations in unit cell volume as a function of temperature and pressure.

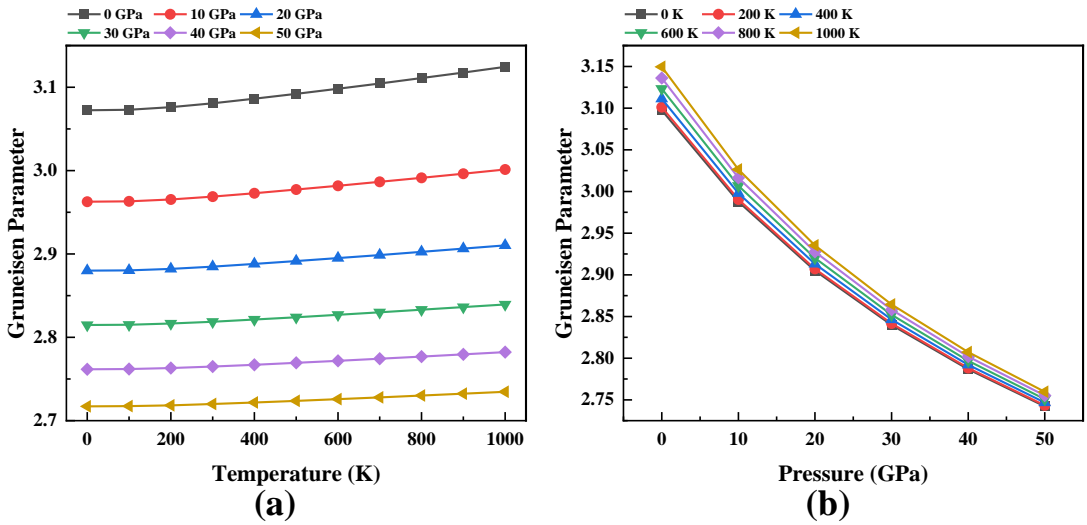

Fig. 5. Variations in Grüneisen parameter as a function of temperature and pressure.

The change in Debye temperature with the variation of temperature and pressure is illustrated in Fig. 6. It is found that the Debye temperature decreases with an increase in temperature while decreasing with an increase in pressure. At 0 GPa and 300 K, the Debye temperature of $Hf_2SeC$ is 640 K which is comparable with its counterpart $Hf_2SC$ (598 K)[51]. We have also computed the Debye temperature ($\Theta_D$) using Anderson's model [88] for both $Hf_2AC$ (A = Se, S). The $\Theta_D$ of $Hf_2SeC$ (640 K, 490 K) is higher than that of $Hf_2SC$ (598 K, 454 K), is in good agreement with the statement that higher $\Theta_D$ corresponds to the harder solids [Section 3.2, hardness of $Hf_2SeC$ is higher than Hf$_2$SeC]. Furthermore, the melting temperature of the titled phase is computed using the model [89] $T_m = 3C_{11} + 1.5C_{33} + 354$, where $C_{11}$ and $C_{33}$ are the elastic constants. The calculated melting point of 1674 K is comparable (slightly lower) with that of Hf$_2$SC (1778 K), which implies that it is also suitable for high-temperature applications. The $T_m$ is related to Young's modulus ($Y$), the higher $T_m$ is for the solids with higher $Y$. The obtained values also agree with the $Y$ values (Hf$_2$SeC – 273 GPa and Hf$_2$SC – 295 GPa). Moreover, the minimum thermal conductivity is also calculated using the formula [90]: $K_{min} = k_B v_m \left(\frac{M}{n\rho N_A}\right)^{-2/3}$ where $k_B$, $v_m$, $N_A$, and $\rho$ are Boltzmann constant, average phonon velocity, Avogadro's number, and density of crystal, respectively. The obtained $K_{min}$ is 0.86 (W/mK) which is almost equal to that Hf$_2$SC [0.85 (W/mK)] [35], smaller than Zr$_2$SeC [1.33 (W/mK)] and M$_2$SC (M = Zr, Nb) [35], also suggests for use in high temperature technology as a coating materials.

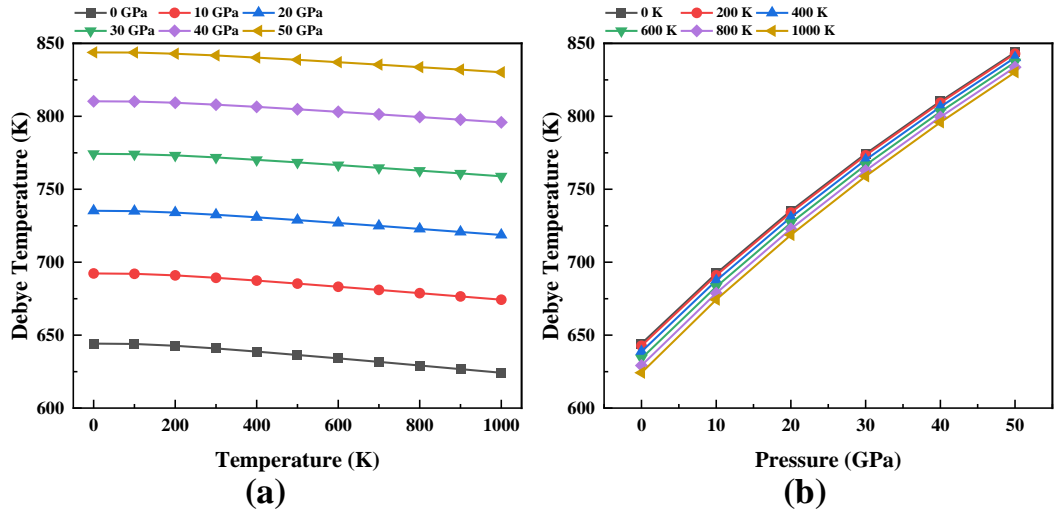

Fig. 6. Variations in Debye temperatureas a function of temperature and pressure.

In Fig. 7, the dependence of the thermal expansion coefficient (*TEC*) on the temperature and pressure is given. The *TEC* is constant at 0 K in the pressure range of 0 to 50 GPa. The *TEC* is increase sharply up to the temperature of 400 K, later, it increases at a gradual rate. The increase in *TEC* at 0 GPa is higher as compared to 50 GPa. At any constant temperature, the *TEC* decreases linearly with the increase in pressure. The computed *TEC* for Hf$_2$SeC is 2.66×10$^{-5}$ K$^{-1}$ is comparable with that of Zr$_2$SeC 3.88×10$^{-5}$ K$^{-1}$[51]. The dependence of specific heat at constant volume ($C_v$) on temperature and pressure is presented in Fig. 8. The variation in $C_v$ as a function of temperature obeys the Debye $T^3$-law and finally becomes constant to reach the classical Dulong-petit (DP) limit i.e., $C_v = 3nNk_B = 195.7$ Jmol$^{-1}$K$^{-1}$. At 0 K, $C_v$ remains constant while decreaseing linearly at higher temperatures as a function of pressure until a temperature where $C_v$ gets converged to reach the DP limit.

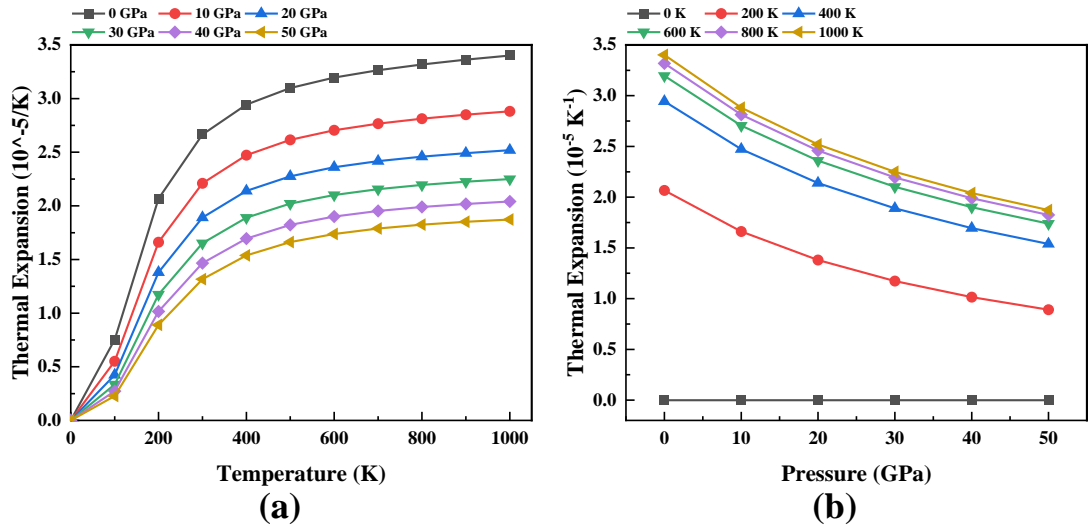

Fig. 7. Variations in thermal expansion coefficient as a function of temperature and pressure.

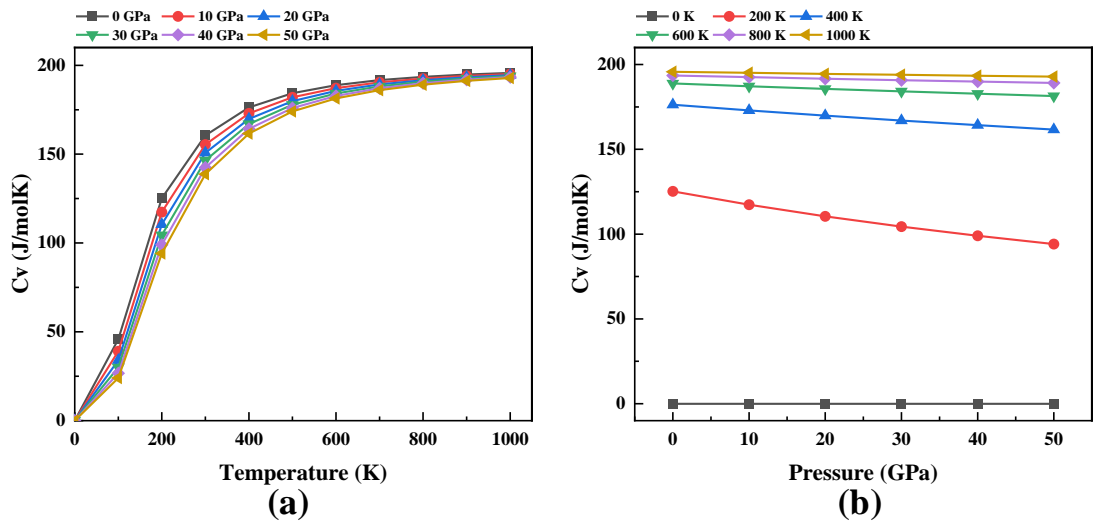

Fig. 8. Variations in specific heat at constant volume as a function of temperature and pressure.

We conclude this section as follows: a good relationship between mechanical and thermal properties is found such as higher Debye temperature for Hf$_2$SeC than Hf$_2$SC due to higher hardness and higher melting point for Hf$_2$SC than Hf$_2$SeC due to higher Young's modulus owing to their interdependence nature. In comparison with TBC materials [91,92], the Debye temperature, melting temperature as well as minimum thermal conductivity values suggest that Hf$_2$SeC is like to be a possible thermal barrier coating material.

## 4. Conclusions

A recently synthesized chalcogenide MAX phase Hf$_2$SeC has been studied using the DFT-based first-principles study. The obtained lattice constants agree well with the previous results. The synthesized titled MAX phase is also mechanically stable. The unequal values of $C_{11}$ and $C_{33}$ imply the anisotropic nature. The obtained parameters for mechanical behavior are comparable with those of other chalcogenide MAX phases. Elastic constants and moduli of Hf$_2$SeC are lower than that of its precursor phase Hf$_2$SC while the hardness follows the reverse trend. The titled boride is like to be 2$^{nd}$ hardest chalcogenide MAX phase. The Hf$_2$SeC is brittle like other MAX phases whereas Hf$_2$SeC is more brittle compared to its precursor Hf$_2$SC. The variation of hardness among Hf$_2$AC (A = Se, S) is well explained in terms of DOSs. The band structure confirms the metallic nature whereas the charge density mapping discloses the mixture of covalent and ionic bonds. The electronic conductivity is anisotropic in nature. Different anisotropic indices, and 2D and 3D plots of $Y$, $B$, $G$, and $v$ also revealed the anisotropic behavior of the mechanical properties. The variation of the thermal properties with temperature and pressure follows the usual trend. The estimated Debye temperature, melting point, and minimum thermal conductivity reveal the possibility of Hf$_2$SeC to be used in high-temperature technology as a TBC material. It is expected that the present study will stimulate the scientific community to discover more Se-based MAX phases in the future.

**Conflicts of interest**

The authors declare no conflicts of interest.

**Funding:** This research did not receive any specific grant from funding agencies in the public, commercial, or not-for-profit sectors.

## References

[1] K. Chen, X. Bai, X. Mu, P. Yan, N. Qiu, Y. Li, J. Zhou, Y. Song, Y. Zhang, S. Du, Z. Chai, Q. Huang, MAX phase Zr2SeC and its thermal conduction behavior, J. Eur. Ceram. Soc. 41 (2021) 4447–4451. https://doi.org/10.1016/j.jeurceramsoc.2021.03.013.

[2]     X. Wang, K. Chen, E. Wu, Y. Zhang, H. Ding, N. Qiu, Y. Song, S. Du, Z. Chai, Q. Huang, Synthesis and thermal expansion of chalcogenide MAX phase Hf2SeC, J. Eur. Ceram. Soc. 42 (2022) 2084–2088. https://doi.org/10.1016/j.jeurceramsoc.2021.12.062.

[3]     M. Sokol, V. Natu, S. Kota, M.W. Barsoum, On the Chemical Diversity of the MAX Phases, Trends Chem. 1 (2019) 210–223. https://doi.org/10.1016/j.trechm.2019.02.016.

[4]     S. Kuchida, T. Muranaka, K. Kawashima, K. Inoue, M. Yoshikawa, J. Akimitsu, Superconductivity in Lu2SnC, Phys. C Supercond. 494 (2013) 77–79. https://doi.org/10.1016/j.physc.2013.04.050.

[5]     T. Rackl, D. Johrendt, The MAX phase borides Zr2SB and Hf2SB, Solid State Sci. 106 (2020) 106316. https://doi.org/10.1016/j.solidstatesciences.2020.106316.

[6]     T. Rackl, L. Eisenburger, R. Niklaus, D. Johrendt, Syntheses and physical properties of the MAX phase boride Nb2SB and the solid solutions N b2 S Bx C1-x(x=0-1), Phys. Rev. Mater. 3 (2019) 054001. https://doi.org/10.1103/PhysRevMaterials.3.054001.

[7]     H. Ding, Y. Li, J. Lu, K. Luo, K. Chen, M. Li, P.O.Å. Persson, L. Hultman, P. Eklund, S. Du, Z. Huang, Z. Chai, H. Wang, P. Huang, Q. Huang, Synthesis of MAX phases Nb2CuC and Ti2(Al0.1Cu0.9)N by A-site replacement reaction in molten salts, Mater. Res. Lett. 7 (2019) 510–516. https://doi.org/10.1080/21663831.2019.1672822.

[8]     L. Mian, L. You-Bing, L. Kan, L. Jun, E. Per, P. Per, R. Johanna, H. Lars, D. Shi-Yu, H. Zheng-Ren, H. Qing, Synthesis of Novel MAX Phase Ti 3 ZnC 2 via A-site-element-substitution Approach, J. Inorg. Mater. 34 (2019) 60. https://doi.org/10.15541/jim20180377.

[9]     M. Li, J. Lu, K. Luo, Y. Li, K. Chang, K. Chen, J. Zhou, J. Rosen, L. Hultman, P. Eklund, P.O.Å. Persson, S. Du, Z. Chai, Z. Huang, Q. Huang, Element Replacement Approach by Reaction with Lewis Acidic Molten Salts to Synthesize Nanolaminated MAX Phases and MXenes, J. Am. Chem. Soc. 141 (2019) 4730–4737. https://doi.org/10.1021/jacs.9b00574.

[10]    H. Fashandi, M. Dahlqvist, J. Lu, J. Palisaitis, S.I. Simak, I.A. Abrikosov, J. Rosen, L. Hultman, M. Andersson, A. Lloyd Spetz, P. Eklund, Synthesis of Ti3AuC2, Ti3Au2C2 and Ti3IrC2 by noble metal substitution reaction in Ti3SiC2 for high-temperature-stable Ohmic contacts to SiC, Nat. Mater. 16 (2017) 814–818. https://doi.org/10.1038/nmat4896.

[11]    Y. Li, Y. Qin, K. Chen, L. Chen, X. Zhang, H. Ding, M. Li, Y. Zhang, S. Du, Z. Chai, Q. Huang, Molten Salt Synthesis of Nanolaminated Sc2SnC MAX Phase, Wuji Cailiao Xuebao/Journal Inorg. Mater. 36 (2021) 773–778. https://doi.org/10.15541/jim20200529.

[12]    B. Tunca, T. Lapauw, O.M. Karakulina, M. Batuk, T. Cabioc'h, J. Hadermann, R. Delville, K. Lambrinou, J. Vleugels, Synthesis of MAX Phases in the Zr-Ti-Al-C System, Inorg. Chem. 56 (2017) 3489–3498. https://doi.org/10.1021/acs.inorgchem.6b03057.

[13]    T. Lapauw, B. Tunca, D. Potashnikov, A. Pesach, O. Ozeri, J. Vleugels, K. Lambrinou, The double solid solution (Zr, Nb)2(Al, Sn)C MAX phase: a steric stability approach, Sci. Rep. 8 (2018) 12801. https://doi.org/10.1038/s41598-018-31271-2.

[14] B. Tunca, T. Lapauw, R. Delville, D.R. Neuville, L. Hennet, D. Thiaudière, T. Ouisse, J. Hadermann, J. Vleugels, K. Lambrinou, Synthesis and Characterization of Double Solid Solution (Zr,Ti) 2 (Al,Sn)C MAX Phase Ceramics, Inorg. Chem. 58 (2019) 6669–6683. https://doi.org/10.1021/acs.inorgchem.9b00065.

[15] M. Griseri, B. Tunca, S. Huang, M. Dahlqvist, J. Rosén, J. Lu, P.O.Å. Persson, L. Popescu, J. Vleugels, K. Lambrinou, Ta-based 413 and 211 MAX phase solid solutions with Hf and Nb, J. Eur. Ceram. Soc. 40 (2020) 1829–1838. https://doi.org/10.1016/j.jeurceramsoc.2019.12.052.

[16] E.Z.S.G. Christopoulos, N. Ni, D.C. Par, D. Horlait, M.E. Fitzpatrick, A. Chroneos, W.E. Lee, Experimental synthesis and density functional theory investigation of radiation tolerance of Zr 3 ( Al 1-x Si x ) C 2 MAX phases, (2017) 1–11. https://doi.org/10.1111/jace.14742.

[17] R. Pan, J. Zhu, Y. Liu, Synthesis, microstructure and properties of (Ti 1− x , Mo x ) 2 AlC phases, Mater. Sci. Technol. 34 (2018) 1064–1069. https://doi.org/10.1080/02670836.2017.1419614.

[18] C. Zuo, C. Zhong, Screen the elastic and thermodynamic properties of MAX solid solution using DFT procedue: Case study on (Ti1-xVx)2AlC, Mater. Chem. Phys. (2020) 123059. https://doi.org/10.1016/j.matchemphys.2020.123059.

[19] M.A. Ali, S.H. Naqib, Recently synthesized (Ti 1−x Mo x ) 2 AlC ($0 \leq x \leq 0.20$) solid solutions: deciphering the structural, electronic, mechanical and thermodynamic properties via ab initio simulations, RSC Adv. 10 (2020) 31535–31546. https://doi.org/10.1039/D0RA06435A.

[20] C.M. Hamm, J.D. Bocarsly, G. Seward, U.I. Kramm, C.S. Birkel, Non-conventional synthesis and magnetic properties of MAX phases (Cr/Mn)2AlC and (Cr/Fe)2AlC, J. Mater. Chem. C. 5 (2017). https://doi.org/10.1039/c7tc00112f.

[21] M.A. Ali, Newly Synthesized Ta- Based MAX Phase (Ta 1− x Hf x ) 4 AlC 3 and (Ta 1− x Hf x ) 4 Al 0.5 Sn 0.5 C 3 ($0 \leq x \leq 0.25$) Solid Solutions: Unravelling the Mechanical, Electronic, and Thermodynamic Properties, Phys. Status Solidi. 258 (2021) 2000307. https://doi.org/10.1002/pssb.202000307.

[22] M.A. Ali, M.R. Khatun, N. Jahan, M.M. Hossain, Comparative study of Mo $_2$ Ga $_2$ C with superconducting *MAX* phase Mo $_2$ GaC: First-principles calculations, Chinese Phys. B. 26 (2017) 033102. https://doi.org/10.1088/1674-1056/26/3/033102.

[23] C. Hu, C.-C. Lai, Q. Tao, J. Lu, J. Halim, L. Sun, J. Zhang, J. Yang, B. Anasori, J. Wang, Y. Sakka, L. Hultman, P. Eklund, J. Rosen, M.W. Barsoum, Mo 2 Ga 2 C: a new ternary nanolaminated carbide, Chem. Commun. 51 (2015) 6560–6563. https://doi.org/10.1039/C5CC00980D.

[24] H. He, S. Jin, G. Fan, L. Wang, Q. Hu, A. Zhou, Synthesis mechanisms and thermal stability of ternary carbide Mo2Ga2C, Ceram. Int. 44 (2018) 22289–22296. https://doi.org/10.1016/j.ceramint.2018.08.353.

[25] H. Chen, D. Yang, Q. Zhang, S. Jin, L. Guo, J. Deng, X. Li, X. Chen, A Series of MAX Phases with MA-Triangular-Prism Bilayers and Elastic Properties, Angew. Chemie - Int. Ed. 58 (2019) 4576–4580. https://doi.org/10.1002/anie.201814128.

[26] Q. Tao, J. Lu, M. Dahlqvist, A. Mockute, S. Calder, A. Petruhins, R. Meshkian, O. Rivin, D. Potashnikov, E.N. Caspi, H. Shaked, A. Hoser, C. Opagiste, R.-M. Galera, R. Salikhov, U. Wiedwald, C. Ritter, A.R. Wildes, B. Johansson, L. Hultman, M. Farle, M.W. Barsoum, J. Rosen, Atomically Layered and Ordered Rare-Earth i -MAX Phases: A New Class of Magnetic Quaternary Compounds, Chem. Mater. 31 (2019) 2476–2485. https://doi.org/10.1021/acs.chemmater.8b05298.

[27] L. Chen, M. Dahlqvist, T. Lapauw, B. Tunca, F. Wang, J. Lu, R. Meshkian, K. Lambrinou, B. Blanpain, J. Vleugels, J. Rosen, Theoretical Prediction and Synthesis of (Cr 2/3 Zr 1/3 ) 2 AlC i -MAX Phase, Inorg. Chem. 57 (2018) 6237–6244. https://doi.org/10.1021/acs.inorgchem.8b00021.

[28] N. Miao, J. Wang, Y. Gong, J. Wu, H. Niu, S. Wang, K. Li, A.R. Oganov, T. Tada, H. Hosono, Computational Prediction of Boron-Based MAX Phases and MXene Derivatives, Chem. Mater. 32 (2020) 6947–6957. https://doi.org/10.1021/acs.chemmater.0c02139.

[29] J. Wang, T.N. Ye, Y. Gong, J. Wu, N. Miao, T. Tada, H. Hosono, Discovery of hexagonal ternary phase Ti2InB2 and its evolution to layered boride TiB, Nat. Commun. 10 (2019) 1–8. https://doi.org/10.1038/s41467-019-10297-8.

[30] M.A. Ali, M.M. Hossain, M.M. Uddin, A.K.M.A. Islam, D. Jana, S.H. Naqib, DFT insights into new B-containing 212 MAX phases: Hf2AB2 (A = In, Sn), J. Alloys Compd. 860 (2021) 158408. https://doi.org/10.1016/j.jallcom.2020.158408.

[31] M.A. Ali, M.M. Hossain, M.M. Uddin, A.K.M.A. Islam, S.H. Naqib, Understanding the improvement of thermo-mechanical and optical properties of 212 MAX phase borides Zr2AB2 (A = In, Tl), J. Mater. Res. Technol. 15 (2021) 2227–2241. https://doi.org/10.1016/j.jmrt.2021.09.042.

[32] M.A. Ali, M.M. Hossain, A.K.M.A. Islam, S.H. Naqib, Ternary boride Hf3PB4: Insights into the physical properties of the hardest possible boride MAX phase, J. Alloys Compd. 857 (2021) 158264. https://doi.org/10.1016/j.jallcom.2020.158264.

[33] M.W. Qureshi, M.A. Ali, X. Ma, Screen the thermomechanical and optical properties of the new ductile 314 MAX phase boride Zr3CdB4: A DFT insight, J. Alloys Compd. 877 (2021) 160248. https://doi.org/10.1016/j.jallcom.2021.160248.

[34] P. Chakraborty, A. Chakrabarty, A. Dutta, T. Saha-Dasgupta, Soft MAX phases with boron substitution: A computational prediction, Phys. Rev. Mater. 2 (2018) 103605. https://doi.org/10.1103/PhysRevMaterials.2.103605.

[35] M.A. Ali, M.M. Hossain, M.M. Uddin, M.A. Hossain, A.K.M.A. Islam, S.H. Naqib, Physical properties of new MAX phase borides M 2 SB (M = Zr, Hf and Nb) in comparison with conventional MAX phase carbides M 2 SC (M = Zr, Hf and Nb): Comprehensive insights, J. Mater. Res. Technol. 11 (2021) 1000–1018.

https://doi.org/10.1016/j.jmrt.2021.01.068.

[36] M. Khazaei, M. Arai, T. Sasaki, M. Estili, Y. Sakka, Trends in electronic structures and structural properties of MAX phases: a first-principles study on M 2 AlC (M = Sc, Ti, Cr, Zr, Nb, Mo, Hf, or Ta), M 2 AlN, and hypothetical M 2 AlB phases, J. Phys. Condens. Matter. 26 (2014) 505503. https://doi.org/10.1088/0953-8984/26/50/505503.

[37] G. Surucu, B. Yildiz, A. Erkisi, X. Wang, O. Surucu, The investigation of electronic, anisotropic elastic and lattice dynamical properties of MAB phase nanolaminated ternary borides: M2AlB2 (M=Mn, Fe and Co) under spin effects, J. Alloys Compd. 838 (2020). https://doi.org/10.1016/j.jallcom.2020.155436.

[38] A. Gencer, G. Surucu, Electronic and lattice dynamical properties of Ti 2 SiB MAX phase, Mater. Res. Express. 5 (2018) 076303. https://doi.org/10.1088/2053-1591/aace7f.

[39] G. Surucu, Investigation of structural, electronic, anisotropic elastic, and lattice dynamical properties of MAX phases borides: An Ab-initio study on hypothetical M2AB (M = Ti, Zr, Hf; A = Al, Ga, In) compounds, Mater. Chem. Phys. 203 (2018) 106–117. https://doi.org/10.1016/j.matchemphys.2017.09.050.

[40] M. Barsoum, T. El-Raghy, The MAX Phases: Unique New Carbide and Nitride Materials, Am. Sci. 89 (2001) 334. https://doi.org/10.1511/2001.4.334.

[41] G. Qing-He, X. Zhi-Jun, T. Ling, Z. Xianjun, J. Guozhu, D. An, L. Rong-Feng, G. Yun-Dong, Y. Ze-Jin, Evidence of the stability of Mo2TiAlC2 from first principles calculations and its thermodynamical and optical properties, Comput. Mater. Sci. 118 (2016) 77–86. https://doi.org/10.1016/J.COMMATSCI.2016.03.010.

[42] M.W. Barsoum, MAX phases: Properties of machinable ternary carbides and nitrides, Wiley-VCH Verlag GmbH & Co. KGaA, Weinheim, Germany, 2013. https://doi.org/10.1002/9783527654581.

[43] M.R. Khatun, M.A. Ali, F. Parvin, A.K.M.A. Islam, Elastic, thermodynamic and optical behavior of V 2 A C ( A = Al, Ga) MAX phases, Results Phys. 7 (2017) 3634–3639. https://doi.org/10.1016/j.rinp.2017.09.043.

[44] A.S. Ingason, M. Dahlqvist, J. Rosen, Magnetic MAX phases from theory and experiments; A review, J. Phys. Condens. Matter. (2016). https://doi.org/10.1088/0953-8984/28/43/433003.

[45] A. Bouhemadou, R. Khenata, Structural, electronic and elastic properties of M2SC (M=Ti, Zr, Hf) compounds, Phys. Lett. A. 372 (2008) 6448–6452. https://doi.org/10.1016/j.physleta.2008.08.066.

[46] I.R. Shein, A.L. Ivanovskii, Elastic properties of superconducting MAX phases from first-principles calculations, Phys. Status Solidi. 248 (2011) 228–232. https://doi.org/10.1002/pssb.201046163.

[47] E. Karaca, P. Byrne, P. Hasnip, H.M. Tutuncu, M. Probert, Electron-Phonon Interaction and Superconductivity in Hexagonal Ternary Carbides Nb2AC (A: Al, S, Ge, As and Sn),

Electron. Struct. (2021). https://doi.org/10.1088/2516-1075/ac2c94.

[48]  M.T. Nasir, M.A. Hadi, S.H. Naqib, F. Parvin, A.K.M.A. Islam, M. Roknuzzaman, M.S. Ali, Zirconium metal-based MAX phases Zr 2 AC (A = Al, Si, P and S): A first-principles study, 28 (2014) 1–16. https://doi.org/10.1142/S0217979215500228.

[49]  M.T. Nasir, A.K.M.A. Islam, MAX phases Nb2AC (A = S, Sn): An ab initio study, Comput. Mater. Sci. 65 (2012) 365–371. https://doi.org/10.1016/J.COMMATSCI.2012.08.003.

[50]  M.F. Cover, O. Warschkow, M.M.M. Bilek, D.R. McKenzie, A comprehensive survey of M 2 AX phase elastic properties, J. Phys. Condens. Matter. 21 (2009) 305403. https://doi.org/10.1088/0953-8984/21/30/305403.

[51]  M.A. Ali, M.W. Qureshi, Newly synthesized MAX phase Zr2SeC: DFT insights into physical properties towards possible applications, RSC Adv. 11 (2021) 16892–16905. https://doi.org/10.1039/d1ra02345d.

[52]  W.-K. Pang, I.M. Low, Understanding and improving the thermal stability of layered ternary carbides and nitrides, in: Adv. Ceram. Matrix Compos., Elsevier, 2018: pp. 429–460. https://doi.org/10.1016/B978-0-08-102166-8.00018-9.

[53]  Y. Fu, B. Wang, Y. Teng, X. Zhu, X. Feng, M. Yan, P. Korzhavyi, W. Sun, The role of group III, IV elements in Nb 4 AC 3 MAX phases (A = Al, Si, Ga, Ge) and the unusual anisotropic behavior of the electronic and optical properties, Phys. Chem. Chem. Phys. 19 (2017) 15471–15483. https://doi.org/10.1039/C7CP01375B.

[54]  M.M. Ali, M.A. Hadi, I. Ahmed, A.F.M.Y. Haider, A.K.M.. Islam, Physical properties of a novel boron-based ternary compound Ti2InB2, Mater. Today Commun. 25 (2020) 101600. https://doi.org/10.1016/j.mtcomm.2020.101600.

[55]  M.A. Hadi, M. Dahlqvist, S.-R.G. Christopoulos, S.H. Naqib, A. Chroneos, A.K.M.A. Islam, Chemically stable new MAX phase V 2 SnC: a damage and radiation tolerant TBC material, RSC Adv. 10 (2020) 43783–43798. https://doi.org/10.1039/D0RA07730E.

[56]  M.D. Segall, P.J.D. Lindan, M.J. Probert, C.J. Pickard, P.J. Hasnip, S.J. Clark, M.C. Payne, First-principles simulation: ideas, illustrations and the CASTEP code, J. Phys. Condens. Matter. 14 (2002) 2717–2744. https://doi.org/10.1088/0953-8984/14/11/301.

[57]  S.J. Clark, M.D. Segall, C.J. Pickard, P.J. Hasnip, M.I.J. Probert, K. Refson, M.C. Payne, First principles methods using CASTEP, Zeitschrift Für Krist. - Cryst. Mater. 220 (2005). https://doi.org/10.1524/zkri.220.5.567.65075.

[58]  J.P. Perdew, K. Burke, M. Ernzerhof, Generalized Gradient Approximation Made Simple, Phys. Rev. Lett. 77 (1996) 3865–3868. https://doi.org/10.1103/PhysRevLett.77.3865.

[59]  H.J. Monkhorst, J.D. Pack, Special points for Brillouin-zone integrations, Phys. Rev. B. 13 (1976) 5188–5192. https://doi.org/10.1103/PhysRevB.13.5188.

[60]  M.W. Barsoum, The MN+1AXN phases: A new class of solids, Prog. Solid State Chem.

28 (2000) 201–281. https://doi.org/10.1016/S0079-6786(00)00006-6.

[61] Y.X. Wang, Z.X. Yan, W. Liu, G.L. Zhou, Structure stability, mechanical properties and thermal conductivity of the new hexagonal ternary phase Ti2InB2 under pressure, Philos. Mag. 100 (2020) 2054–2067. https://doi.org/10.1080/14786435.2020.1754485.

[62] Y. Medkour, A. Bouhemadou, A. Roumili, Structural and electronic properties of M2InC (M = Ti, Zr, and Hf), Solid State Commun. 148 (2008) 459–463. https://doi.org/10.1016/j.ssc.2008.09.006.

[63] M.W. Qureshi, X. Ma, G. Tang, R. Paudel, Ab initio predictions of structure and physical properties of the Zr2GaC and Hf2GaC MAX- phases under pressure, Sci. Reports |. 11 (2021) 1–23. https://doi.org/10.1038/s41598-021-82402-1.

[64] M. Born, On the stability of crystal lattices. I, Math. Proc. Cambridge Philos. Soc. 36 (1940) 160–172. https://doi.org/10.1017/S0305004100017138.

[65] F. Mouhat, F.-X. Coudert, Necessary and sufficient elastic stability conditions in various crystal systems, Phys. Rev. B. 90 (2014) 224104. https://doi.org/10.1103/PhysRevB.90.224104.

[66] R. Hill, The Elastic Behaviour of a Crystalline Aggregate, Proc. Phys. Soc. Sect. A. 65 (1952) 349–354. https://doi.org/10.1088/0370-1298/65/5/307.

[67] W. Voigt, Lehrbuch der Kristallphysik, Vieweg+Teubner Verlag, Wiesbaden, 1966. https://doi.org/10.1007/978-3-663-15884-4.

[68] A. Reuss, Berechnung der Fließgrenze von Mischkristallen auf Grund der Plastizitätsbedingung für Einkristalle ., ZAMM - J. Appl. Math. Mech. / Zeitschrift Für Angew. Math. Und Mech. 9 (1929) 49–58. https://doi.org/10.1002/zamm.19290090104.

[69] X.-Q. Chen, H. Niu, D. Li, Y. Li, Modeling hardness of polycrystalline materials and bulk metallic glasses, Intermetallics. 19 (2011) 1275–1281. https://doi.org/10.1016/j.intermet.2011.03.026.

[70] N. Miao, B. Sa, J. Zhou, Z. Sun, Theoretical investigation on the transition-metal borides with Ta3B4-type structure: A class of hard and refractory materials, Comput. Mater. Sci. 50 (2011) 1559–1566. https://doi.org/10.1016/J.COMMATSCI.2010.12.015.

[71] Y. Pan, X. Wang, S. Li, Y. Li, M. Wen, DFT prediction of a novel molybdenum tetraboride superhard material, RSC Adv. 8 (2018) 18008–18015. https://doi.org/10.1039/c8ra02324g.

[72] S.F. Pugh, XCII. Relations between the elastic moduli and the plastic properties of polycrystalline pure metals, London, Edinburgh, Dublin Philos. Mag. J. Sci. 45 (1954) 823–843. https://doi.org/10.1080/14786440808520496.

[73] M. Roknuzzaman, M.A. Hadi, M.A. Ali, M.M. Hossain, N. Jahan, M.M. Uddin, J.A. Alarco, K. Ostrikov, First hafnium-based MAX phase in the 312 family, Hf 3 AlC 2 : A first-principles study, J. Alloys Compd. 727 (2017) 616–626.

https://doi.org/10.1016/j.jallcom.2017.08.151.

[74] D.G. Pettifor, Theoretical predictions of structure and related properties of intermetallics, Mater. Sci. Technol. 8 (1992) 345–349. https://doi.org/10.1179/026708392790170801.

[75] M.A. Ali, M.M. Hossain, M.A. Hossain, M.T. Nasir, M.M. Uddin, M.Z. Hasan, A.K.M.A. Islam, S.H. Naqib, Recently synthesized (Zr 1-x Ti x ) 2 AlC (0 ≤ x ≤ 1) solid solutions: Theoretical study of the effects of M mixing on physical properties, J. Alloys Compd. 743 (2018) 146–154. https://doi.org/10.1016/j.jallcom.2018.01.396.

[76] Y. Zhou, Z. Sun, Electronic structure and bonding properties of layered machinable and ceramics, Phys. Rev. B - Condens. Matter Mater. Phys. 61 (2000) 12570–12573. https://doi.org/10.1103/PhysRevB.61.12570.

[77] J. Wang, Y. Zhou, Dependence of elastic stiffness on electronic band structure of nanolaminate M 2 AlC ( M = Ti , V , Nb , and Cr ) ceramics, 214111 (2004) 1–9. https://doi.org/10.1103/PhysRevB.69.214111.

[78] Y.Y. Liu, Y.X. Wang, Z.X. Yan, W. Liu, G.L. Zhou, K.Z. Xiong, J.B. Gu, Pressure effects on structure, mechanical properties and thermal conductivity of V2SnC: a first-principles study, Philos. Mag. 0 (2021) 1–16. https://doi.org/10.1080/14786435.2021.1988747.

[79] M.W. Qureshi, X. Ma, G. Tang, R. Paudel, Ab initio predictions of structure and physical properties Hf 2 GaC MAX of the - Zr 2 GaC and - phases under pressure, (2021) 1–24. https://doi.org/10.1038/s41598-021-82402-1.

[80] W. Sailuam, I. Fongkaew, S. Limpijumnong, K. Phacheerak, A first principles investigation on the structural, elastic, and mechanical properties of MAX phase M3AlC2 (M= Ta, Ti, V) as a function of pressure, Comput. Condens. Matter. 30 (2022) e00638. https://doi.org/10.1016/j.cocom.2021.e00638.

[81] H.M. Ledbetter, Elastic properties of zinc: A compilation and a review, J. Phys. Chem. Ref. Data. 6 (1977) 1181–1203. https://doi.org/10.1063/1.555564.

[82] A.K.M.A. Islam, A.S. Sikder, F.N. Islam, NbB2: a density functional study, Phys. Lett. A. 350 (2006) 288–292. https://doi.org/10.1016/j.physleta.2005.09.085.

[83] J. Wang, Y. Zhou, T. Liao, Z. Lin, First-principles prediction of low shear-strain resistance of Al3BC3: A metal borocarbide containing short linear BC2 units, Appl. Phys. Lett. 89 (2006) 021917. https://doi.org/10.1063/1.2220549.

[84] S.I. Ranganathan, M. Ostoja-Starzewski, Universal Elastic Anisotropy Index, Phys. Rev. Lett. 101 (2008) 055504. https://doi.org/10.1103/PhysRevLett.101.055504.

[85] J. Gonzalez-Julian, Processing of MAX phases: From synthesis to applications, J. Am. Ceram. Soc. 104 (2021) 659–690. https://doi.org/10.1111/jace.17544.

[86] A. Otero-De-La-Roza, D. Abbasi-Pérez, V. Luaña, Gibbs2: A new version of the quasiharmonic model code. II. Models for solid-state thermodynamics, features and

implementation, Comput. Phys. Commun. 182 (2011) 2232–2248. https://doi.org/10.1016/j.cpc.2011.05.009.

[87] M.A. Blanco, E. Francisco, V. Luaña, GIBBS: Isothermal-isobaric thermodynamics of solids from energy curves using a quasi-harmonic Debye model, Comput. Phys. Commun. 158 (2004) 57–72. https://doi.org/10.1016/j.comphy.2003.12.001.

[88] O.L. Anderson, A simplified method for calculating the debye temperature from elastic constants, J. Phys. Chem. Solids. 24 (1963) 909–917. https://doi.org/10.1016/0022-3697(63)90067-2.

[89] M.E. Fine, L.D. Brown, H.L. Marcus, Elastic constants versus melting temperature in metals, Scr. Metall. 18 (1984) 951–956. https://doi.org/10.1016/0036-9748(84)90267-9.

[90] G.A. Slack, The Thermal Conductivity of Nonmetallic Crystals, Solid State Phys. - Adv. Res. Appl. 34 (1979) 1–71. https://doi.org/10.1016/S0081-1947(08)60359-8.

[91] Y. Zhou, X. Lu, H. Xiang, Z. Feng, Preparation, mechanical, and thermal properties of a promising thermal barrier material: Y4Al2O9, J. Adv. Ceram. 4 (2015) 94–102. https://doi.org/10.1007/s40145-015-0141-5.

[92] Y. Zhou, H. Xiang, X. Lu, Z. Feng, Z. Li, Theoretical prediction on mechanical and thermal properties of a promising thermal barrier material: Y4Al2O9, J. Adv. Ceram. 4 (2015) 83–93. https://doi.org/10.1007/s40145-015-0140-6.